\documentclass[prl,twocolumn,showpacs,amsmath,amssymb]{revtex4}
\usepackage{graphicx}% Include figure files
\usepackage{dcolumn}% Align table columns on decimal point
\usepackage{bm}% bold math
\usepackage{color}
\newcommand{\nn}{\nonumber}

\begin{document}

\preprint{APS/123-QED}

\title{Textures of Spin-Orbit Coupled $F\!=\!2$ Spinor Bose Einstein Condensates}% Force line breaks with \\

\author{Takuto Kawakami}
% \email{Second.Author@institution.edu}
\author{Takeshi Mizushima}
\author{Kazushige Machida}
\affiliation{Department of Physics, Okayama University, Okayama 700-8530 Japan}%

\date{\today}

\begin{abstract}
We study the textures of $F\!=\! 2$ spinor Bose{-}Einstein condensates (BECs) with spin-orbit coupling (SOC) induced by a synthetic {non-Abelian} gauge field.
On the basis of the analysis of the SOC energy and the numerical calculation of the Gross-Pitaevskii equation, 
{we demonstrate that the textures originate from the helical modulation of the order parameter (OP) due to the SOC.} In particular, the cyclic OP consists of {two-dimensional} lattice textures, such as the hexagonal lattice and the {$\frac{1}{3}$}-vortex lattice, commonly understandable as the {two-dimensional} network of the helical modulations.
\end{abstract}

\pacs{03.75.Lm, % Tunneling, Josephson effect, Bose-Einstein condensates in periodic potentials, solitons, vortices, and topological excitations (see also 74.50.+r Tunneling phenomena; Josephson effects in superconductivity)
      03.75.Mn, % Multicomponent condensates; spinor condensates
      67.85.Fg, % Multicomponent condensates; spinor condensates
      67.85.Jk}  % Other Bose-Einstein condensation phenomena

\maketitle
Motivated by {the} recent {successful} {generation of} gauge fields in neutral cold atoms~\cite{NIST1,NIST2},
much attention has been paid to Bose-Einstein condensates (BECs) in Abelian~{\cite{NIST1}} {and} non-Abelian~{\cite{gauge, juzeliunas, NIST2}} gauge fields.
{The synthetic field technique {has allowed} access {to} the quantum Hall regime~\cite{fetter} and {has realized} nontrivial textures in spinor BECs~\cite{pietila,ho, wang, yzhang}}.

{It is well known in fermion systems such as topological insulators {and} superconductors 
that spin-orbit coupling (SOC) plays a crucial role.
This is particularly true for fermionic superfluid $^3$He because SOC due originally to the $^3$He
nuclear dipole-dipole interaction is decisive {in determining} the order parameter (OP) textures{,} although its 
force is so tiny {that it modifies} the Cooper pair structure itself{,} as first pointed out by Leggett~\cite{leggett}.
Since in bosonic superfluids, namely the spinor BEC~{\cite{F=1, F=2}}, no corresponding ``natural'' SOC {exists},
one may implement it by using {a} ``synthetic'' gauge potential {to better} understand and control the OP space in spinor BEC systems.
In fact{,} we see that by introducing SOC{,} they exhibit a plethora of 
types of the OP textures in a controlled way,
such as fractional vortices, {two-dimensional} periodic textures, {and} skyrmions.}

{The pseudospin states are the eigenenergy states of the laser{-}atom interaction, sufficiently isolated from the other states.
The adiabatic motion of {these} states produces a vector potential in real space~\cite{gauge, juzeliunas, NIST2, wilczek}.}
{{Several} schemes to generate gauge fields {are} suggested.}
One of them is the $\Lambda$ scheme~\cite{juzeliunas,gauge}, {which} 
is useful to realize a symmetric spin-orbit coupling term, such as the Rashba{-}type SOC.
{The different scheme {is} implemented by the NIST group~\cite{NIST2},
{where the SOC consists of the equal contribution of the Rashba and Dresselhaus types.}}

{T}he gauge transformation cannot remove the {spatially uniform} non-Abelian vector potential.
Thus, {the} BECs {have} {a} spatially modulated OP and textures {due to finite momentum}~\cite{ho, wang, yzhang}.
{Nontrivial textures due to} the Rashba{-}type SOC in pseudospin $F\!=\!1/2$ and $F\!=\!1$ BECs have been studied recent{ly}~\cite{wang}.
The {two} possible textures are {the {one-dimensional}} plane and standing waves{;} {in the former, the {phase} of the {OP} varies and in the latter, the amplitude oscillates along a favorable direction.}
{In $F\!=\!1$ BECs,}
{the stable region of the plane and standing waves} is equivalent to that of the ferromagnetic and polar phases, respectively~\cite{F=1}.
However, we emphasize {here} that {an} open question remains of how textural structures can emerge in $F\!=\! 2$ spinor BECs, where the cyclic phase distinct from the manifold of the ferromagnetic and polar phase can be the magnetic ground state.

{The aim in this {Rapid Communication} is to clarify the role of the SOC in bosonic superfluids.}
We focus on $F\!=\!2$ spinor BECs, since they have magnetic ground states {that include} the cyclic phase. On the basis of the analysis of the SOC energy, we reveal that the plane (standing) wave realized 
in ferromagnetic (polar) phase can be interpreted as the rotation of the OP in pseudospin space{,} 
which simultaneously propagates along one direction in real space. 
{In addition, {skyrmions of the uniaxial polar OP} emerge in the small size of the system.}
This analysis enables us to understand the emergence of nontrivial textural structures in the cyclic phase. 
{We} demonstrate that {the hexagonal lattice and {$\frac{1}{3}$}-vortex lattice states} are energetically competitive. 
We also calculate the phase diagram where the spin-spin interaction favors the cyclic phase.

% ==========================================================================================================
% ==========================================================================================================
We consider a zero temperature $F\!=\!2$ spinor BEC with a SOC.
The single{-}particle Hamiltonian including the Rashba-like SOC term is described as
\begin{eqnarray}\label{SOC}
	&&H_\mathrm{0} = \int d^2\bm{r} \vec{\Psi}^\dagger(\bm{r})\left\{ h_0 -\frac{\hbar^2\kappa}{\sqrt{2}m}\hat{\bm{M}}_\mathrm{SO}\cdot\bm{\nabla}\right\}\vec{\Psi}(\bm{r}), \\
&&\hat{\bm{M}}_\mathrm{SO} = 
\left[\begin{array}{ccccc}
	0 & \bm{e}_- &0  &0 & 0 \\
	\bm{e}_+ & 0 & \beta\bm{e}_- &0 &0 \\
	0 &\beta\bm{e}_+ &0 & \beta\bm{e}_- &0  \\
	0 &0  & \beta\bm{e}_+ & 0 &\bm{e}_-\\
 	0 &0  &0 & \bm{e}_+ &0
\end{array}\right], \nn
\end{eqnarray}
where $\vec{\Psi}(\bm{r}) = [\psi_2,\ \psi_1,\ \psi_0,\ \psi_{-1},\ \psi_{-2}]^T$ is the OP vector in {an} $F\!=\!2$ BEC, $h_0\!=\! -\nabla^2/2m + V_\mathrm{pot}(r)$ consists of the kinetic energy and the trap potential term, and $\bm{e}_{\pm}=\hat{\bm{x}} \pm i\hat{\bm{y}}$.
The planar hexapod {setup} introduced by Juzeli\=unas {\it et al.}~\cite{juzeliunas} realizes Rashba-like SOC with $\beta\!=\!1$. 
However, in an $F=2$ spinor BEC system, the experimental {setup} for realizing the precise Rashba-type SOC {form} with $\beta\!=\!\sqrt{6}/2$ 
has not been found {thus} far.

To reveal the role of the SOC term in Eq.~(\ref{SOC}),
we {first simplify the OP as
\begin{eqnarray}\label{psiR}
	\vec{\Psi}(\bm{r})=\hat{R}(\bm{k}\cdot{\bm r},\hat{\bm{n}}_\mathrm{R})\vec{\Psi}_\mathrm{I}(\bm{r}_0),
\end{eqnarray}
where $\hat{R}(\theta,\hat{\bm n})$ denotes the rotation matrix with the angle $\theta$ about $\hat{\bm{n}}$} in the pseudospin space{;} 
$\hat{\bm n}_\mathrm{R}$ and $\hat{\bm k}$ are the rotation axis and the modulation vector in the {$x$-$y$} plane, respectively.
$\vec{\Psi}_\mathrm{I}$ is an arbitrary OP vector in the pseudospin space at a certain point {$\bm{r}_0$}.
{By substituting Eq.~(\ref{psiR}) {in Eq.} (\ref{SOC}), one can obtain the SOC energy {density}{:}}
\begin{eqnarray}\label{ARSOE}
&&h_\mathrm{SO}(\bm{k},\hat{\bm{n}}_\mathrm{R},\vec{\Psi}_I)=-\frac{\hbar^2\kappa}{16\sqrt{2}m}\vec{\Psi}^{\prime\dagger}_{\mathrm{I}}\hat{\bm{M}}'_\mathrm{SO}\cdot\bm{k}\vec{\Psi}'_\mathrm{I} \\
&& \hat{\bm{M}}'_\mathrm{SO} = 
	{\hat{\bm{n}}_\mathrm{R}}
	\left[\begin{array}{ccccc}
	4A &0 &0&0& 0\\
	0 &4 & 0 & 0&0\\
	\sqrt{6}\omega^{2}C & 0 & 0 & 0 & \sqrt{6}\omega^{-2}C\\
	0 & 0 & 0& 4 & 0\\
	0& 0 &0&0 & 4A
	\end{array}\right] \nn \\ \nn
&&	+ i{\hat{\bm{n}}_\mathrm{R}\times\hat{\bm z}}\left[\begin{array}{ccccc}
	0 &-i\omega^{-1}A &0&\omega^{3}C& 0\\
	2\omega A &0 & 0 & 0&2i\omega^{-3}C\\
	0 & \omega B & 0 & \omega^{-1}B & 0\\
	2\omega^{3}C & 0 & 0& 0 & 2\omega^{-1}A\\
	0& \omega^{-3}C &0& \omega^{-1}A & 0
	\end{array}\right], \nn
\end{eqnarray}
where $A\!=\!(1+\sqrt{6}\beta)$, $B\!=\!\sqrt{3/2}(3+\sqrt{6}\beta)$, $C\!=\!(3-\sqrt{6}\beta)$, $\omega\!=\!\exp[i\bm{k}\cdot\bm{r}]$, 
$\vec{\Psi}_\mathrm{I}'\!=\!\hat{R}(\pi/3, \hat{\bm{e}}_{111})\vec{\Psi}_\mathrm{I}$, and $\hat{\bm{e}}_{111}\!\parallel\!\hat{\bm{x}}+\hat{\bm{y}}+\hat{\bm{z}}$.

{The diagonal elements of {$\hat{\bm M}'_\mathrm{SO}$ in Eq.~(\ref{ARSOE})} are found to be energetically dominant, compared to the offdiagonal elements.}
{This is because the} oscillation terms with $\omega^n$ ($n\neq0$) hardly contribute {to energetics} when they are integrated {over $\bm{r}$}.
The dominant {elements} {{are} proportional to ${\bm k}\cdot{\bm n}_\mathrm{R}$}. 
{(i) The SOC favors the situation {where} the rotation axis $\hat{\bm n}_R$ corresponds to the modulation vector $\hat{\bm k}$}.
We call this spatial modulation the {{\it helical modulation}}.
In addition{,} we define the angular resolved SOC energy density $h_\mathrm{AR}(\bm{k},\vec{\Psi}_\mathrm{I})\!\equiv\!h_\mathrm{SO}(\bm{k}, \hat{\bm n}_\mathrm{R}\!=\!\hat{\bm k}, \vec{\Psi}_\mathrm{I})$, which denotes the SOC energy density for $\vec{\Psi}_\mathrm{I}$ with a given modulation vector $\bm{k}$.
{(ii) The helical modulation of the OP is independent of the details of SOC term, such as the precise Rashba-type SOC ($\beta=\sqrt{6}/2$, that is, $C=0$) and the Rashba-like SOC ($\beta=1$) implemented by the {planar} hexapod {setup}~\cite{juzeliunas}.} 

Here, we assume that the atoms in pseudospin states interact through the most symmetric {interaction{:}}
\begin{eqnarray}\label{Hint}
	H_\mathrm{int} = \frac{1}{2}\int d^2\bm{r} \left[c_0 n^2 + c_1 \bm{S}\cdot\bm{S} + c_2 |A_{00}|^2 \right],
\end{eqnarray}
where {$n \!=\! \vec{\Psi}^\dagger\vec{\Psi}$, $\bm{S}\!=\!\vec{\Psi}^\dagger\bm{\sigma}\vec{\Psi}$, and $A_{00} \!=\! (2\psi_2\psi_{\!-\!2} - 2\psi_1\psi_{\!-\!1} + \psi_0\psi_0)/\sqrt{5}$ are the {particle} density, the spin {density}, and the singlet pair amplitude, respectively.}
{The purpose of using this assumption is to} 
concentrate on the symmetry breaking due to the SOC term.
In addition, {it is known that the minimization of $H_\mathrm{int}$ leads to four magnetic ground states:}
Ferromagnetic (FM), biaxial polar (BP), uniaxial polar (UP), and cyclic (CY) phase~\cite{F=2}.

\begin{figure}[tb]
		\begin{center}
			\includegraphics[width=70mm]{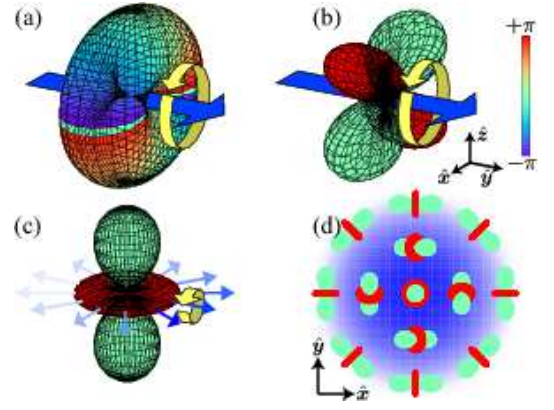}
			\caption{
					(Color online)
					OP profiles in pseudospin space and energetically favored modulation vectors of ground states{:} (a) FM, (b) BP, {and} (c) UP.
					The color on the surface corresponds to the phase of OP.
					{The UP-skyrmion texture is shown in (d).}}
			\label{fig:rotation}
		\end{center}
	\end{figure}

We numerically minimize the full Gross-Pitaevskii energy functional $H_0+H_\mathrm{int}$ without any restriction by using the imaginary time evolution scheme in the presence of the cylindrical symmetric trap.
{The numerical results presented in this Rapid Communication are obtained by setting the parameter $\beta\!=\!$1 and all of the results are not changed qualitatively by the parameter $\beta$.}
The numerical calculation reveals that for the {FM} parameter region, as shown in Fig.~\ref{fig:rotation}(a), the resulting OP can be simplified to the {one-dimensional} helical modulation of the uniform FM OP along the $\hat{\bm y}$ axis, that is{,} {$\vec{\Psi}({\bm r}) \!=\! \hat{R}(ky,\hat{\bm y}) \vec{\Psi}_{\rm FM}$} with $\vec{\Psi}_{\rm FM} \!=\! \hat{R}^{-1}(\pi/3,\hat{\bm{e}}_{111})[1,0,0,0,0]^{\rm T}$. 
{The wave number $k$ depends on $\kappa$.} In the same way, the favorable OP for the BP is given by replacing $\vec{\Psi} _{\rm FM}$ {in} $\vec{\Psi} _{\rm BP} \!=\! \hat{R}^{-1}(\pi/3,\hat{\bm{e}}_{111})[1,0,0,0,1]^{\rm T}$, which is shown in Fig.~\ref{fig:rotation}(b). 
Using these simplified OPs, the SOC energy density reduces to 
{$|h_\mathrm{AR}(k\hat{\bm y}, \vec{\Psi}_\mathrm{FM(BP)})| =\hbar^2\kappa k(\sqrt{6}\beta+1)/4\sqrt{2}$}, independent of the FM and BP. These one dimensional helical modulation{s} of the FM and BP {are} consistent {with} those realized in $F\!=\!1/2$ and $F\!=\!1$ systems~\cite{wang}.

In contrast, the favorable OP in the UP state reduces to {$\vec{\Psi}({\bm r}) \!=\! \hat{R}({\bm k}\cdot{\bm r},\hat{\bm k})\vec{\Psi}_\mathrm{UP}$ with $\vec{\Psi}_\mathrm{UP}=[0,0,1,0,0]^{\rm T}$}, where {the modulation vector ${\bm k}$} can be arbitrary in the $x$-$y$ plane
as shown in Fig.~\ref{fig:rotation}(c).
{Because} {${\bm k}$} {is not unique}, the most favorable modulation turns out to be {${\bm k} \!\parallel\! {\bm{r}}$}, resulting in the skyrmion texture in Fig.~\ref{fig:rotation}(d).
The SOC energy density of the UP skyrmion is found to be {$|h_\mathrm{AR}(kr,\vec{\Psi}_\mathrm{UP})| = 3\hbar^2\kappa k(\sqrt{6}\beta + 1)/16\sqrt{2}$}, which is less than that of the FM and BP. 

{We note that} the UP and BP are degenerate in the absence of SOC $\kappa\!=\!0$.
In the {presence} of the Rashba-like SOC, 
the energetics of the UP skyrmion texture and BP standing wave is
determined by the {ratio} of two length scales $1/\kappa$ and the Thomas-Fermi radius {$R_\mathrm{TF}$}{,} 
which characterize the modulation {in real space} and the size of the condensate, respectively.
The stable region of the UP skyrmion is {$1/\kappa \!\gtrsim\! R_\mathrm{TF}$}, while the BP standing wave becomes stable for {$1/\kappa \!\lesssim\! R_\mathrm{TF}$}. 

\begin{figure}[tb]
		\begin{center}
			\includegraphics[width=80mm]{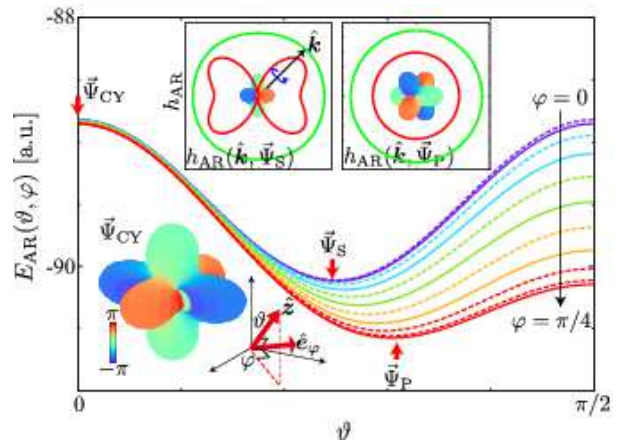}
			\caption{
					(Color online)
					The main panel describes the SOC energy 
					$E_\mathrm{AR}(\vartheta, \varphi)\equiv\int d\phi_{\bm k} h_\mathrm{AR}({\bm k}, \vec{\Psi}_\mathrm{I})$.
					The bottom inner figures show the cyclic OP $\vec{\Psi}_{\rm CY}$ and
					the angles $\vartheta$ and $\varphi${,} which denote 
					the $\hat{\bm{z}}$ axis of $\vec{\Psi}_\mathrm{I}$.
					The two upper panels are the angular resolved SOC energy  
					$|h_\mathrm{AR}(\hat{\bm{k}},\vec{\Psi}_\mathrm{I})|$ for the cases of
					$\vec{\Psi}_{\mathrm{I}}\!=\!\vec{\Psi}_{\mathrm{S}}$ and $\vec{\Psi}_{\mathrm{I}}\!=\!\vec{\Psi}_{\mathrm{P}}$.
					In these panels, {the {radii} of} the outer and inner lines indicate 
					$|h_\mathrm{AR}(\hat{\bm{k}},\vec{\Psi}_\mathrm{I})|$ and 
					the deviation of $|h_\mathrm{AR}(\hat{\bm{k}},\vec{\Psi}_\mathrm{I})|$ from the value at the antinode,
					{in the $k_x$-$k_y$ plane.} 
					}
			\label{fig:SOCterm}
		\end{center}
	\end{figure}

% ==========================================================================================================
% ==========================================================================================================
{Next, we move} to the textures in the CY parameter region with $c_1\!\ge\! 0$ and $c_2\!\ge\! 0$. 
The OP of the CY is characterized by a nodal structure as shown in inner bottom of Fig.~\ref{fig:SOCterm}, which has eight point nodes at $\!\pm\! \hat{\bm x}\!\pm\! \hat{\bm y}\!\pm\! \hat{\bm z}$, $\!\pm\! \hat{\bm x}\!\mp\! \hat{\bm y}\!\pm\! \hat{\bm z}$, $\!\pm\! \hat{\bm x}\!\pm\! \hat{\bm y}\!\mp\! \hat{\bm z}$, and $\!\pm\! \hat{\bm x}\!\mp\! \hat{\bm y}\!\mp\! \hat{\bm z}$ and six antinodes. Their point nodes are connected with saddle lines. 
The inherent difference from {those} shown in Fig.~\ref{fig:rotation} 
{arises from} the three{-}dimensional {form of the nodal points}. 
This three dimensionality gives rise {to} the textures with the two-dimensional modulation vectors as mentioned {below}.

Before going to the numerical results, we discuss the role of the SOC term in Eq.~(\ref{SOC}) for the cyclic phase. It is convenient to introduce two OP vectors $\vec{\Psi}_\mathrm{I}\!=\!\hat{R}(\arccos[1/\sqrt{3}], \hat{\bm{x}}\!+\!\hat{\bm y})\vec{\Psi}_\mathrm{CY}\!\equiv\!\vec{\Psi}_\mathrm{P}$ and $\vec{\Psi}_\mathrm{I}\!=\!\hat{R}(\pi/2, \hat{\bm{x}}\!+\!\hat{\bm y})\vec{\Psi}_\mathrm{CY}\!\equiv\!\vec{\Psi}_\mathrm{S}$. Here, $\vec{\Psi}_\mathrm{CY}\!=\![i/2,0,1/\sqrt{2},0,i/2]$ denotes the simple form of the cyclic OP shown in inner bottom in Fig.~\ref{fig:SOCterm} and $\vec{\Psi}_\mathrm{P}$ ($\vec{\Psi}_\mathrm{S}$) describes the cyclic OP where one of point nodes (saddle points) points to the $\hat{\bm z}$ axis.
The two upper panels in Fig.~\ref{fig:SOCterm} show the angular resolved SOC energy
$|h_\mathrm{AR}(\bm{k},\vec{\Psi}_\mathrm{S})|$ and $|h_\mathrm{AR}(\bm{k},\vec{\Psi}_\mathrm{P})|$ for a given $|{\bm k}|$.
In the case of $\vec{\Psi}_\mathrm{S}$, the OP has four point nodes and two antinodes in the $x$-$y$ plane. It is seen in the upper panels of Fig.~\ref{fig:SOCterm} that for the case of $\vec{\Psi}_\mathrm{S}$, {$|h_\mathrm{AR}(\bm{k},\vec{\Psi}_\mathrm{S})|$} becomes maximum (minimum) when {$\hat{\bm k}$} points {in} the nodal (antinodal) direction. 
In contrast, the {angularly} resolved SOC energy density for $h_\mathrm{AR}(\bm{k},\vec{\Psi}_\mathrm{P})$ becomes cylindrically symmetric in the $x$-$y$ plane.
The main panel of Fig.~\ref{fig:SOCterm} shows {$E_\mathrm{AR}(\vartheta, \varphi)\!\equiv\!\int d\phi_{\bm k} h_\mathrm{AR}({\bm k}, \vec{\Psi}_\mathrm{I})$ with $\vec{\Psi}_I \!=\! \hat{R}(\vartheta, \hat{\bm e}_\varphi)\vec{\Psi}_\mathrm{CY}$ and $\hat{\bm{e}}_\varphi \!=\!-\!\sin\varphi \hat{\bm x}\!+\!\cos \varphi \hat{\bm y}$}. 
Here, it is found that the most stable situation in the cyclic region is $\vec{\Psi}_\mathrm{I}\!=\!\vec{\Psi}_\mathrm{P}$, where one of the point nodes points to the $\hat{\bm z}$ direction. However, note that the SOC in Eq.~(\ref{SOC}) also gives rise to the helical modulation of the OP, which rotates the direction of the point node from {the} $\hat{\bm z}$ axis.

\begin{figure}[tb]
		\begin{center}
			\includegraphics[width=80mm]{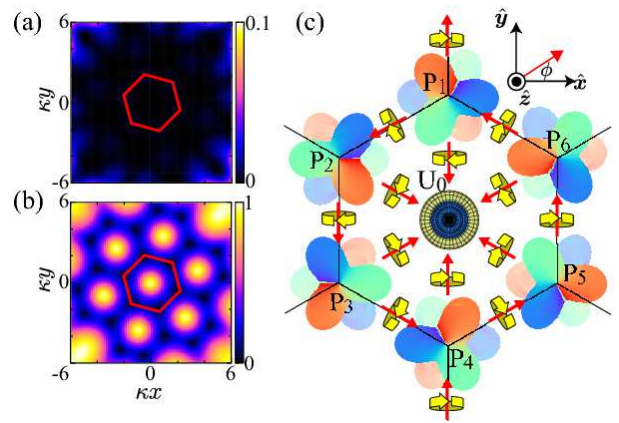}
			\caption{
					(Color online)
					{Profiles of} the CY{-}UP lattice texture: 
					(a) {{pseudospin} density $\bm{S}(\bm{r})$}, 
					(b) singlet amplitude {$A_{00}(\bm{r})$}, and (c) the order parameter. 
					The parameters are {set to be} {$\kappa R_\mathrm{TF}\!=\!10.1$} and $c_1/c_0\!=\!c_2/c_0\!=\!0.2$.}
			\label{fig:UPcore}
		\end{center}
	\end{figure}

\begin{figure}[tb]
		\begin{center}
			\includegraphics[width=80mm]{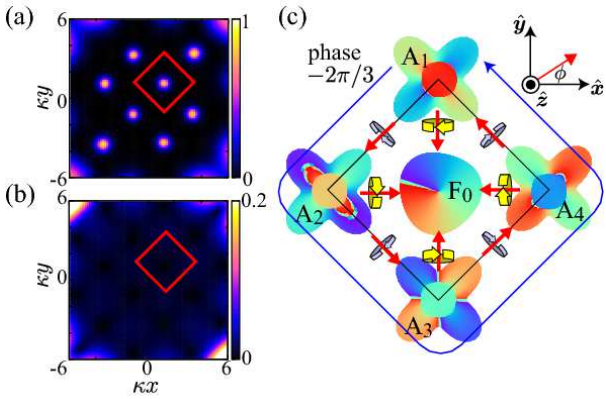}
			\caption{
					(Color online)
					{Profiles of the $\frac{1}{3}$-vortex lattice} texture: 
					(a) {{pseudospin} density $\bm{S}(\bm{r})$}, 
					(b) singlet amplitude {$A_{00}(\bm{r})$}, and (c) the order parameter. 
					The parameters are {set to be} {$\kappa R_\mathrm{TF}\!=\!10.1$}, $c_1/c_0\!=\!0.2$, and $c_2/c_0\!=\!20$.}
			\label{fig:FMcore}
		\end{center}
	\end{figure}

Let us argue the {stable textures in real space for $c_1/c_0\!\ge\!0$ and $c_2/c_0\!\ge\!0$}.  
{Based on the numerical calculation of the full GP equation including the SOC term,} 
we find three kinds of textures: 
The uniform cyclic texture, the CY{-}UP hexagonal lattice texture shown in Fig.~\ref{fig:UPcore}, and the {$\frac{1}{3}$}-vortex lattice texture shown in Fig.~\ref{fig:FMcore}. 

	\begin{figure}[t]
		\begin{center}
			\includegraphics[width=80mm]{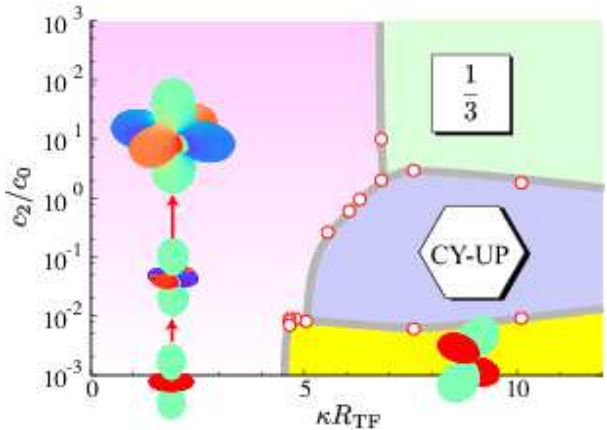}
			\caption{
					(Color online)
					Phase diagram of textures
					spanned by coupling constant of SOC $\kappa$ and that of the spin-spin interaction $c_2/c_0$.
					We fix the parameter $c_1/c_0\!=\!0.2$. 
					The stable phase for $\kappa R_{\rm TF} \!\lesssim\!1$ continuously 
					changes from the UP skyrmion to uniform CY texture. 
					In the $\kappa R_{\rm TF}\!\gtrsim\!1$ region, three textures appear: The BP standing wave 
					for the small $c_2/c_0$, the CY{-}UP lattice ({hexagon}), and {$\frac{1}{3}$}-vortex lattice (square). }
			\label{fig:phase}
		\end{center}
	\end{figure}

The emergence of the CY{-}UP lattice texture is straightforwardly comprehensible, on the basis of the argument of the SOC energy in the cyclic OP. 
As shown in Figs.~\ref{fig:UPcore}(a) and {\ref{fig:UPcore}}(b), {this texture consists of the hexagonal unit cell in which the polar state is localized.}
Here, we {look carefully into} the texture inside the hexagonal unit cell.
{As seen in Fig.~\ref{fig:UPcore}(c), the unit cell consists of the uniaxial polar OP at $U_0$ enclosed by cyclic OPs at $P_1$ to $P_6$, which are obtained from the numerical calculation of the GP equation. At all the points {$P_1 \hbox{-} P_6$}, one of the point nodes always points to the $\hat{\bm z}$ axis, which makes the SOC energy lower, as discussed {in Fig.~\ref{fig:SOCterm}}. In addition, {at $P_1$} the other three point nodes are located at $(\theta,\phi) \!=\! (\theta _0,-\pi/3)$, $(\theta _0,\pi/3)$, and $(\theta _0,\pi)$ with $\theta _0 \!=\! \arccos(2\sqrt{2}/3)$. By propagating the helical modulation around one of the saddle point{s}, the OP $\vec{\Psi}_\mathrm{P}$ at $P_1$ can be continuously transformed to that at $P_2$, which orients the point node $(\theta _0,-\pi/3)$ at $P_1$ to the $\hat{\bm z}$ direction at $P_2$. The point node $(\theta _0,\pi/3)$ and $(\theta_0,\pi)$ at $P_1$ is oriented to the $\hat{\bm z}$ direction at $P_6$ and $P_4$. In the same way, the OP at $P_2$ can propagate with the helical modulation along the circumference $P_2\!\rightarrow\!P_3\rightarrow\!\cdots\rightarrow\! P_1$. Note that the cyclic phase at $P_1$ can have six modulation vectors which have isotropic energy gain in the SOC term{,} as {shown} in Fig.~\ref{fig:SOCterm}. The central region of the unit cell in Fig.~\ref{fig:UPcore}(c) is occupied by the UP OP to avoid frustration of the cyclic OP at $U_0$. Hence, in this sense this network of the helical modulation can be regarded as the close-packed hexagonal lattice of the cyclic OP whose point node points to the $\hat{\bm z}$ axis.}

The other possible texture in the cyclic region is displayed in Fig.~\ref{fig:FMcore}, which can be the ferromagnetic core at $F_0$ enclosed by cyclic OPs. At $A_1$ in Fig.~\ref{fig:FMcore}(c){,} the antinode points to the $\hat{\bm z}$ axis, which cannot be the ground state of the SOC energy in Eq.~(\ref{SOC}). The OP at $A_1$ has fourfold symmetry in the $x$-$y$ plane. The OP at $A_1$ can be continuously transformed to that at {$A_2$}
by the helical modulation {along $-\!\hat{\bm x}\!-\!\hat{\bm y}$} with the rotation angle $\pi/2$.
After the helical modulation along the path {$A_1\!\rightarrow\!A_2\!\rightarrow\!\cdots\! A_1$} {in the same way}, the OP results in the shape where the antinode points to the $\hat{\bm z}$ axis. However, the resulting OP differs from the original one and has the phase shift $+2\pi/3$. In order to recover the single{-}value nature at $A_1$, the phase shift $-2\pi/3$ is compensated by the {$U(1)$} phase of the OP. Hence, this texture with fourfold symmetry can be regarded as the square lattice of the {$\frac{1}{3}$} vortices, in the sense that the {$U(1)$} phase of the cyclic OP continuously changes by $2\pi/3$ along the path $A_1 \!\rightarrow\! A_2 \!\rightarrow \! \cdots \! \rightarrow\! A_1$. Here, the central region at $F_0$ is occupied by the ferromagnetic core because of frustration of the cyclic OP. It is expected that the {$\frac{1}{3}$} vortex also appears in the cyclic phase of $F\!=\! 2$ spinor BEC under rotation~\cite{1/3}, which behaves as the non-Abelian {vortices}. 

In Fig.~\ref{fig:phase}, we summarize the phase diagram spanned by $\kappa R_{\rm TF}$ and $c_2/c_0$. For $\kappa R_{\rm TF}\!\lesssim\!1$, the stable texture continuously changes from the UP skyrmion to the uniform CY. For $\kappa R_{\rm TF}\!\gtrsim\!1$, the standing wave of the BP appears in the small $c_2/c_0$ region. As $c_2/c_0$ increases, the hexagonal lattice of UP core and the {$\frac{1}{3}$}-vortex lattice become energetically competitive. The former texture is the close-packed lattice of the CY OP with the point node pointing to {the} $\hat{\bm z}$ axis. On the other hand, the FM core in the {$\frac{1}{3}$}-vortex lattice relatively favors the {$H_\mathrm{int}$}, which becomes stable in the large $c_2/c_0$ region. 

% ==========================================================================================================
% ==========================================================================================================
In summary, we have studied the stable textures of $F\!=\!2$ spinor BEC {with} a spin-orbit interaction. 
Based on the detailed analysis of the spin-orbit interaction, we find that the ferromagnetic and biaxial polar OPs have a unique direction for the helical modulation, while that of the uniaxial polar OP is not unique. The nonuniqueness leads to the emergence of the skyrmion textures{,} which become stable in the parameter region where the polar phase is stable. Moreover, we have computed the phase diagram where the cyclic phase is the magnetic ground state. The phase diagram is covered by two-dimensional lattices, such as the CY{-}UP lattice and {$\frac{1}{3}$}-vortex lattice. These lattice structures are understandable as the two-dimensional network of the helical modulation of the cyclic order parameters. 
{We emphasize that those novel textures are not seen in rotating $F\!=\!2$ spinor BEC~\cite{pogosov} and appear inherently due to SOC.}

{We should mention that the Hamiltonian for} spinor BECs under synthetic gauge fields does not necessarily have the pseudospin rotation symmetry~\cite{yzhang}, which can be altered by the laser configuration and so on. Hence, the general form of the {$H_\mathrm{int}$} is open for future study. 
Moreover, the experimental observation of the textures remains a challenging problem different from those with hyperfine spins~\cite{stamper-kurn}

{\it Note added:} Recently, we became aware of a preprint by Xu {\it et al.}~\cite{xu}, which has some overlap with our results. 

% ==========================================================================================================
% ==========================================================================================================
The authors thank M. Ichioka for helpful discussions.
This work was supported by JSPS and the ``Topological Quantum Phenomena" KAKENHI on innovation areas from MEXT.

% ==========================================================================================================
% ==========================================================================================================


\begin{thebibliography}{99}

%\if0
\bibitem{NIST1}
%Y.-J. Lin, R. L. Compton, A. R. Perry, W. D. Phillips, J. V. Porto, and I. B. Spielman, Phys. Rev. Lett. {\bf 102}, 130401 (2009).
Y.-J. Lin {\it et al.}, Phys. Rev. Lett. {\bf 102}, 130401 (2009);
%Y.-J. Lin, R. L. Compton, K. Jimenez-Garcia, J. V. Porto, and I. B. Spielman, Nature (London) {\bf 462}, 628 (2009).
Y.-J. Lin {\it et al.}, Nature (London) {\bf 462}, 628 (2009).

\bibitem{NIST2}
%Y.-J. Lin, K. Jim\'enez-Garcia, and I. B. Spielman, Nature (London) {\bf 471}, 83 (2011).
Y.-J. Lin {\it et al.}, Nature (London) {\bf 471}, 83 (2011).

\bibitem{gauge}
%J. Ruseckas {\it et al.}, Phys. Rev. Lett. {\bf 95}, 010404 (2005);
J. Ruseckas, {G. Juzeli\=unas, P. \"{O}hberg, and M. Fleischhauer,} Phys. Rev. Lett. {\bf 95}, 010404 (2005);
%T. D. Stanescu {\it et al.}, Phys. Rev. A {\bf 78}, 023616 (2008).
T. D. Stanescu, {B. Anderson, and V. Galitski}, Phys. Rev. A {\bf 78}, 023616 (2008).

\if0
	\bibitem{ruseckas}
	%J. Ruseckas, G. Juzeli\=unas, P. \"{O}hberg, and M. Fleischhauer, Phys. Rev. Lett. {\bf 95}, 010404 (2005).
	J. Ruseckas {\it et al.}, Phys. Rev. Lett. {\bf 95}, 010404 (2005).
	
	\bibitem{stanescu}
	%T. D. Stanescu, B. Anderson, and V. Galitski, Phys. Rev. A {\bf 78}, 023616 (2008).
	T. D. Stanescu {\it et al.}, Phys. Rev. A {\bf 78}, 023616 (2008).
\fi

\bibitem{juzeliunas}
{G. Juzeli\=unas, J. Ruseckas, and J. Dalibard, Phys. Rev. A {\bf 81}, 053403 (2010).}
%G. Juzeli\=unas {\it et al.}, Phys. Rev. A {\bf 81}, 053403 (2010).


\bibitem{fetter}
A. L. Fetter, Rev. Mod. Phys. {\bf 81}, 647 (2009).

\bibitem{ho}
T.-L. Ho and S. Zhang, arXiv:1007.0650.

\bibitem{wang}\label{wang}
{C. Wang, C. Gao, C.-M Jian, and H. Zhai, Phys. Rev. Lett. {\bf 105}, 160403 (2010).}
%C. Wang {\it et al.}, Phys. Rev. Lett. {\bf 105}, 160403 (2010).

\bibitem{yzhang}
Y. Zhang, L. Mao, and C. Zhang, arXiv:1102.4045.

\bibitem{pietila}
V. Pietil\"{a} and M. M\"{o}tt\"{o}nen, Phys. Rev. Lett. {\bf 102}, 080403 (2009).
%V. Pietil\"{a} {\it et al.}, Phys. Rev. Lett. {\bf 102}, 080403 (2009).

\bibitem{leggett}
A. J. Leggett, Phys. Rev. Lett. {\bf 31}, 352 (1973).

\bibitem{F=1}
{T. Ohmi and K. Machida, J. Phys. Soc. Jpn. {\bf 67}, 1822 (1998);
T.-L. Ho, Phys. Rev. Lett. {\bf 81}, 742 (1998).}

\bibitem{F=2}
{C. V. Ciobanu, S.-K. Yip, and T.-L. Ho, Phys. Rev. A {\bf 61}, 033607 (2000)};
%C. V. Ciobanu {\it et al.}, Phys. Rev. A {\bf 61}, 033607 (2000);
M. Ueda and M. Koashi, {\it ibid.} {\bf 65}, 063602 (2002).

\bibitem{wilczek}
F. Wilczek and A. Zee, Phys. Rev. Lett. {\bf 52}, 2111 (1984).

\bibitem{1/3}
{G. W. Semenoff and F. Zhou, Phys. Rev. Lett. {\bf 98}, 100401 (2007); M. Kobayashi, Y. Kawaguchi, M. Nitta, and M. Ueda, {\it ibid.} {\bf 103}, 115301 (2009); J. A. M. Huhtam\"{a}ki, T. P. Simula, M. Kobayashi, and K. Machida, Phys. Rev. A {\bf 80}, 051601(R) (2009); H. M. Adachi {\it et al.}, J. Phys. Soc. Jpn {\bf 78}, 113301 (2009); {\bf 79}, 044301 (2010).}
%G. W. Semenoff and F. Zhou, Phys. Rev. Lett. {\bf 98}, 100401 (2007); M. Kobayashi {\it et al.}, Phys. Rev. Lett. {\bf 103}, 115301 (2009); J. A. M. Huhtam\"{a}ki {\it et al.}, Phys. Rev. A {\bf 80}, 051601(R) (2009); {H. M. Adachi {\it et al.}, J. Phys. Soc. Jpn {\bf 78}, 113301 (2009);} {{\it ibid} {\bf 79}, 044301 (2010).}

	\if0
	\bibitem{semenoff}
	%G. W. Semenoff and F. Zhou, Phys. Rev. Lett. {\bf 98}, 100401 (2007).
	G. W. Semenoff and F. Zhou, Phys. Rev. Lett. {\bf 98}, 100401 (2007).
	
	\bibitem{kobayashi2}
	%M. Kobayashi, Y. Kawaguchi, M. Nitta, and M. Ueda, Phys. Rev. Lett. {\bf 103}, 115301 (2009).
	M. Kobayashi, Y. Kawaguchi, M. Nitta, and M. Ueda, Phys. Rev. Lett. {\bf 103}, 115301 (2009).
	
	\bibitem{huhtamaki}
	%J. A. M. Huhtam\"{a}ki, T. P. Simula, M. Kobayashi, and K. Machida, Phys. Rev. A {\bf 80}, 051601(R) (2009).
	J. A. M. Huhtam\"{a}ki, T. P. Simula, M. Kobayashi, and K. Machida, Phys. Rev. A {\bf 80}, 051601(R) (2009).
	
	\bibitem{adachi}
	%{H. M. Adachi, Y. Tsutsumi, and K. Machida, J. Phys. Soc. Jpn {\bf 78}, 113301 (2009).}
	{H. M. Adachi, Y. Tsutsumi, and K. Machida, J. Phys. Soc. Jpn {\bf 78}, 113301 (2009).}
	
	\bibitem{adachi2}
	%{H. M. Adachi, Y. Tsutsumi, and K. Machida, J. Phys. Soc. Jpn {\bf 79}, 044301 (2010).}
	{H. M. Adachi, Y. Tsutsumi, and K. Machida, J. Phys. Soc. Jpn {\bf 79}, 044301 (2010).}
	\fi
\bibitem{pogosov}
{W. V. Pogosov, R. Kawate, T. Mizushima, and K. Machida, Phys. Rev. Lett. {\bf 72}, 063605 (2005).}
%{W. V. Pogosov {\it et al.}, Phys. Rev. A {\bf 72}, 063605 (2005).}

\bibitem{stamper-kurn}
%L. E. Sadler, J. M. Higbie, S. R. Leslie, M. Vengalattore, and D. M. Stamper-Kurn, Nature (London) 443, 312 (2006).
{L. E. Sadler {\it et al.}, Nature (London) {\bf 443}, 312 (2006).}

\bibitem{xu}
Z. F. Xu, R. L\"{u}, and L. You, Phys. Rev. A {\bf 83} {053602} (2011).
%\fi
\end{thebibliography}
\end{document}